\documentclass[preprint,aps,showkeys]{revtex4}
\usepackage{graphicx}

\begin{document}

\title{Dynamical Evolution of Photons in Plasma Waves}
\author{Zhigang Bu$^{1)}$}
\author{Yuee Luo$^{1)}$}
\author{Hehe Li$^{1)}$}
\author{Wenbo Chen$^{1)}$}
\author{Peiyong Ji$^{1)2)}$}
\email{pyji@staff.shu.edu.cn} \affiliation{1) Department of
Physics, Shanghai University, Shanghai 200444, China\\
2) The Shanghai Key Lab of Astrophysics, Shanghai  200234, China}
\date{\today}

\begin{abstract}
On the viewpoint of corpuscular model an electromagnetic radiation can be regarded as a system composed of photons with different energies and momenta, which provides us a method being different from the Maxwell wave theory to describe the interaction of electromagnetic waves with plasmas. In this paper the evolution behavior of a single photon and the collective effect of a photon system in plasma waves are uniformly described in the frame of photon dynamics. In a small-amplitude plasma wave the modulation of photon dynamical behavior by the plasma wave can be treated as perturbation, and the photon acceleration effect and photon Landau damping are investigated in the linear theory. In a plasma wave with arbitrary amplitude the photon evolution trajectories in phase space and coordinate space are obtained by solving the dynamical equations, and the trapping condition and possibility of photons in the given plasma wave are also discussed.
\end{abstract}
\maketitle

\section{Introduction}
The interaction of intense fields with plasmas is an important hotspot issue in plasma physics, which is involved in many area, for example the excitation of wakefield, laser-plasma acceleration, fast ignition fusion[1-7], etc. A laser pulse can excite a plasma wave via the effect of ponderomotive force when propagating through a plasma, and motions of electrons and photons in plasmas are also modulated by this plasma wave. In the interaction process the laser-plasma system can be described by the Maxwell-fluid theory which has achieved huge success in the last two decades[8-16].

On the viewpoint of corpuscular theory an electromagnetic field interacting with plasmas can be treated as a nonequilibrium photon system governed by the kinetic equation, which is called as photon kinetic theory[17-19]. Based on the photon kinetic theory the dynamical behavior of a single photon in a background plasma can be extended to the evolution of the entire photon system. As a special photon system, laser pulse has a lot of advantages compared with other electromagnetic radiations, such as good directivity, high intensity and narrow spectrum width, etc. We can treat the laser pulse as an ordered photon current described by the kinetic theory. The connection between the photon system and a single photon is the photon number distribution in phase space which is essentially the statistical weight of photons with various modes in the entire photon system. The evolution of an electromagnetic radiation in plasmas can be derived by making weighted average to all photon behaviors. Due to the photon kinetic theory being a very effective method for describing the propagation properties of electromagnetic fields in plasmas, the research on single photon dynamics is worth trying.

There are some similarities between photon and electron behaviors in plasma waves. For instance, if the photon velocity matches with the phase velocity of the plasma wave, the photon can be accelerated effectively by the plasma wave[20-23] and the photon system can produce Landau collisionless damping to the plasma wave[24-26], etc., and so do the electrons[27-31]. However the differences are obvious. The photon is not influenced by the electrostatic field of the plasma wave because of the electric neutrality, whereas the electron is. The propagation behavior of photons in plasmas is related to the refractive index, i.e., electron number density of the plasma. The refractive index of a plasma can be equivalent to a special potential field acting on photons, and photons will get some equivalent forces when moving in this potential field just as electrons are affected by the electrostatic forces in plasma waves. In this sense, the dynamical evolution of a single photon in a plasma is determined by the electron number density of the plasma. In inhomogeneous plasma the equivalent forces acting on photons depend on their positions in the plasma, and the photons with different modes are also influenced by different forces owing to dispersion effects.

The purpose of this paper is to investigate the propagation and evolution behaviors of photons in a background plasma wave by means of the photon dynamical theory. The photon dynamical equations are expressed in Hamiltonian formulation based on the approximation of geometrical optics[22], and the photon frequency is regarded as Hamiltonian for a photon. Based on the dynamical theory we describe some phenomena in the interaction process of photons with plasma wave in a unified frame. If the amplitude of a plasma wave is small enough, the modulation of the photon dynamical behavior by the plasma wave can be treated as perturbation. In Sec. 2 the photon dynamical equations are treated by the perturbation theory, and the photon acceleration effect, along with the collective collisionless damping originated from the laser pulse, is analyzed to the second order. In Sec. 3 the dynamical evolution of a single photon in a plasma wave with arbitrary amplitude is investigated. The photon evolution trajectory in phase space is derived by solving photon dynamical equations. The trapping condition and possibility of photons in a given plasma wave are also analysed, and the motion equations in coordinate space for both trapped and untrapped photons are analytically obtained. At last we give a discussion in Sec. 4.

\section{Photon Behavior in Small-Amplitude Plasma Wave}
Let us consider one dimensional model for simplicity, and suppose that the electromagnetic wave propagates along $x$ axis. In this paper we assume that the intensity of the electromagnetic wave is not strong enough to change the configuration of the plasma wave prominently so that the reaction from the electromagnetic wave (or the photon system) to the plasma wave can be ignored. When an electromagnetic wave propagates in a plasma wave, the disturbed electron number density modulates the frequency $\omega$ and wave vector $k$ of the electromagnetic wave to produce frequency shifting, which can be described by the Maxwell wave theory. Nevertheless, we will discuss this problem in an alternative way. By considering the electromagnetic wave as a system of photon gas, let us discuss the kinetic evolution of a single photon in a plasma wave. If the amplitude of a plasma wave is small, $\delta n\ll n_{0}$, where $\delta n$ is the disturbed electron number density and $n_{0}$ is the ambient plasma density, the photon kinetics can be expressed by perturbation expansion:
\begin{equation}
x=x_{0}+v_{0}t+x_{1}+x_{2}+\cdots\cdots,
\end{equation}
\begin{equation}
k=k_{0}+k_{1}+k_{2}+\cdots\cdots,
\end{equation}
\begin{equation}
\omega=\omega_{0}+\omega_{1}+\omega_{2}+\cdots\cdots,
\end{equation}
\begin{equation}
v=v_{0}+v_{1}+v_{2}+\cdots\cdots,
\end{equation}
where $x$ and $v$ are the position and velocity of the photon, $\omega$ and $k$ are regarded as the energy and momentum of the photon. The subscript 0 denotes the initial quantity, and 1 and 2 denote the first- and second-order corrections respectively, and $v_{0}=c^{2}k_{0}/\omega_{0}$.

Assuming that the configuration of the plasma wave can be expressed by a function of $f(x,t)=f(k_{pl}x-\omega_{pl}t)$, where $k_{pl}$ and $\omega_{pl}$ are the wave number and frequency of the plasma wave. The Hamiltonian of the photon is proportional to its frequency $\omega$ as determined by the dispersion relation of the electromagnetic wave in the plasma
\begin{equation}
H=\hbar\omega=\hbar\sqrt{c^{2}k^{2}+\omega_{p}^{2}},
\end{equation}
where $\omega_{p}$ is the local plasma frequency determined by the disturbed electron number density in the plasma wave as $\omega_{p}^{2}=\omega_{p0}^{2}(1+f(x,t))$. The photon evolution in a plasma wave is governed by the dynamical equations in Hamiltonian formulation
\begin{equation}
v=\frac{dx}{dt}=\frac{\partial H}{\partial
k},
\end{equation}
\begin{equation}
\frac{dk}{dt}=-\frac{\partial H}{\partial x}.
\end{equation}
For simplicity, we suppose the plasma wave has the shape of cosine, $f=\kappa\cos(k_{pl}x-\omega_{pl}t)$, where $\kappa=\delta n/n_{0}$ is the amplitude of the plasma wave and a small quantity $\kappa\ll1$. Dynamical equation (7) gives
\begin{equation}
\frac{dk}{dt}=\frac{\omega_{p0}^{2}k_{pl}\kappa}{2\omega}\sin(k_{pl}x-\omega_{pl}t).
\end{equation}
Substituting Eqs.(1)-(3) into Eq.(8) the first- and second-order equations for the photon momentum are derived as
\begin{equation}
\frac{dk_{1}}{dt}=\frac{\omega_{p0}^{2}k_{pl}\kappa}{2\omega_{0}}\sin\left(k_{pl}x_{0}-\Omega t\right),
\end{equation}
\begin{eqnarray}
\frac{dk_{2}}{dt}=&&\frac{\omega_{p0}^{2}k_{pl}\kappa}{2\omega_{0}}\left[k_{pl}x_{1}\cos\left(k_{pl}x_{0}-\Omega t\right)-\frac{c^{2}k_{0}k_{1}}{\omega_{0}^{2}}\sin\left(k_{pl}x_{0}-\Omega t\right)\right.\nonumber\\
&&~~~~~~~~~~~~\left.-\frac{\omega_{p0}^{2}\kappa}{2\omega_{0}^{2}}\cos\left(k_{pl}x_{0}-\Omega t\right)\sin\left(k_{pl}x_{0}-\Omega t\right)\right],
\end{eqnarray}
in which $\Omega=\omega_{pl}-k_{pl}v_{0}$. The integral of Eq.(9) gives a first-order correction of the photon momentum in the small-amplitude plasma wave
\begin{equation}
k_{1}=\frac{\omega_{p0}^{2}k_{pl}\kappa}{2\omega_{0}\Omega}\left[\cos\left(k_{pl}x_{0}-\Omega t\right)-\cos\left(k_{pl}x_{0}\right)\right].
\end{equation}

In plasma waves the photon frequency is modulated by the electron number density, which is known as the photon acceleration effect. By using of the dispersion relation of an electromagnetic field in plasmas we can expand the photon frequency to second-order as
\begin{eqnarray}
\omega&=&\omega_{0}\left\{1+\frac{c^{2}k_{0}^{2}}{2\omega_{0}^{2}}\left(\frac{2k_{1}}{k_{0}}+\frac{\omega_{p0}^{2}\kappa}
{c^{2}k_{0}^{2}}\cos\left(k_{pl}x_{0}-\Omega t\right)\right)\right.\nonumber\\
&&+\left.\frac{c^{2}k_{0}^{2}}{2\omega_{0}^{2}}\left[\left(\frac{1}{k_{0}^{2}}-\frac{c^{2}}{\omega_{0}^{2}}\right)k_{1}^{2}+\frac{2k_{2}}{k_{0}}
-\frac{\omega_{p0}^{2}k_{pl}\kappa}{c^{2}k_{0}^{2}}x_{1}\sin\left(k_{pl}x_{0}-\Omega t\right)\right.\right.\nonumber\\
&&\left.\left.-\frac{\omega_{p0}^{4}\kappa^{2}}{4c^{2}k_{0}^{2}\omega_{0}^{2}}\cos^{2}\left(k_{pl}x_{0}-\Omega t\right)-\frac{\omega_{p0}^{2}\kappa
k_{1}}{k_{0}\omega_{0}^{2}}\cos\left(k_{pl}x_{0}-\Omega t\right)\right]\right\},
\end{eqnarray}
where $\omega_{0}$ and $k_{0}$ are the initial photon frequency and momentum respectively. The frequency shift of photon in the plasma wave can be analyzed by using of Eq.(12). Defining the frequency shift at time $t$ as $\Delta\omega=\omega(t)-\omega(0)$, where $\omega(0)$ is the initial photon frequency, and using the first-order equation of photon momentum (11), it is given that
\begin{equation}
\Delta\omega=-\frac{\omega_{p0}^{2}\omega_{pl}\kappa}{2\omega_{0}\Delta vk_{pl}}\left[\cos\left(k_{pl}x_{0}+k_{pl}\Delta vt\right)-\cos\left(k_{pl}x_{0}\right)\right]
\end{equation}
when calculated to the linear term, where the parameter $\Delta v=v_{0}-v_{\phi}$, and $v_{\phi}=\omega_{pl}/k_{pl}$ denotes the phase velocity of the plasma wave. If the photon velocity $v_{0}$ is very close to the phase velocity of the plasma wave, i.e., in the limit of $\Delta v\rightarrow0$, we get the maximum frequency shift
\begin{equation}
\lim_{\Delta v\rightarrow0}\Delta\omega=\frac{k_{pl}t\omega_{p0}^{2}v_{\phi}\kappa}{2\omega_{0}}\sin\left(k_{pl}x_{0}\right),
\end{equation}
which is consistent with the earlier publications[21].

On the viewpoint of corpuscular property an electromagnetic radiation can be regarded as a particle system composed of photons with various modes. The evolution behavior of entire electromagnetic radiation in plasmas can be derived by making statistical summation of all photon behaviors in theory. Now we will discuss energy exchanging of the electromagnetic wave with the plasma wave based on the photon dynamics, which is a kind of collective effect of photon system known as photon Landau damping.

Taking the derivative of Eq.(12) with respect to time $t$, the average rate of energy change of the photon to second-order in plasma wave is given by
\begin{eqnarray}
\left\langle\frac{d\omega}{dt}\right\rangle&=&\frac{c^{2}}{\omega_{0}}\left(1-\frac{c^{2}k_{0}^{2}}{\omega_{0}}\right)\langle
k_{1}\dot{k}_{1}\rangle+\frac{c^{2}k_{0}}{\omega_{0}}\langle\dot{k}_{2}\rangle\nonumber\\
&&-\frac{\omega_{p0}^{2}k_{pl}\kappa}{2\omega_{0}}\left(\langle\dot{x}_{1}\sin\left(k_{pl}x_{0}-\Omega
t\right)\rangle-\Omega\langle x_{1}\cos\left(k_{pl}x_{0}-\Omega t\right)\rangle\right)\nonumber\\
&&-\frac{c^{2}k_{0}\omega_{p0}^{2}\kappa}{2\omega_{0}^{3}}\left(\langle\dot{k}_{1}\cos\left(k_{pl}x_{0}-\Omega
t\right)+\Omega\langle k_{1}\sin\left(k_{pl}x_{0}-\Omega t\right)\rangle\right).
\end{eqnarray}
Here the dot stands for the derivative with respect to the time, and the bracket $\langle\cdots\rangle$ denotes the average over initial position $x_{0}$ in one wavelength range of the plasma wave. The first-order correction of photon velocity can be obtained from dynamical equation Eq.(6) as
\begin{equation}
v_{1}=\dot{x}_{1}
=\frac{c^{2}\omega_{p0}^{2}\kappa}{2\omega_{0}^{3}}\left(\frac{\omega_{p0}^{2}k_{pl}}{\omega_{0}\Omega}-k_{0}\right)\cos\left(k_{pl}x_{0}-\Omega
t\right)-\frac{c^{2}\omega_{p0}^{4}k_{pl}\kappa}{2\omega_{0}^{4}\Omega}\cos(k_{pl}x_{0}).
\end{equation}
The integral of Eq.(16) gives
\begin{equation}
x_{1}=\frac{c^{2}\omega_{p0}^{2}\kappa}{2\omega_{0}^{3}\Omega}\left(k_{0}-\frac{\omega_{p0}^{2}k_{pl}}{\omega_{0}\Omega}\right)\left[\sin\left(k_{pl}x_{0}-\Omega
t\right)-\sin\left(k_{pl}x_{0}\right)\right]-\frac{c^{2}\omega_{p0}^{4}k_{pl}\kappa}{2\omega_{0}^{4}\Omega}t\cos\left(k_{pl}x_{0}\right).
\end{equation}
Substituting Eqs.(9)-(11), (16) and (17) into Eq.(15) we get
\begin{equation}
\left\langle\frac{d\omega}{dt}\right\rangle=-\frac{c^{2}\omega_{p0}^{6}\omega_{pl}k_{pl}^{2}\kappa^{2}}{8\omega_{0}^{5}}\frac{\partial}{\partial\Omega}\left(\frac{1}{\Omega}\sin(\Omega
t)\right)-\frac{c^{2}\omega_{p0}^{4}k_{0}\omega_{pl}k_{pl}\kappa^{2}}{4\omega_{0}^{4}}\frac{1}{\Omega}\sin(\Omega t).
\end{equation}

In order to extend the rate of energy change of the photon to the whole electromagnetic field we need to computer the sum of all photons in the field. Supposing that the initial photon number distribution is $N(\omega_{0})$, the average rate of energy change of the electromagnetic field is given by
\begin{equation}
\overline{\langle\dot{\omega}\rangle}=\int\langle\dot{\omega}\rangle
N\left(\omega_{0}\right)d\omega_{0}.
\end{equation}
Eq.(18) indicates that $\langle\dot{\omega}\rangle$ is oscillatory and damped rapidly, excepting at the resonance singularity $\Omega=0$. Thus the main contribution in Eq.(19) comes from the area nearby the singularity, which can be worked out by considering the long time limit and expressed in terms of a delta function
\begin{equation}
\lim_{t\rightarrow\infty}\frac{1}{\Omega}\sin\left(\Omega
t\right)=\pi\delta\left(\Omega\right)=\frac{\pi\omega_{\phi}^{2}k_{\phi}}{\omega_{p0}^{2}k_{pl}}\delta\left(\omega_{0}-\omega_{\phi}\right),
\end{equation}
where $\omega_{\phi}$ is resonance frequency of the photon with the plasma wave as defined by $v_{0}(\omega_{0}=\omega_{\phi})=v_{\phi}$. Since $\partial/\partial\Omega=-\left(\omega_{0}^{2}k_{0}\right)/\left(\omega_{p0}^{2}k_{pl}\right)\partial/\partial \omega_{0}$, we obtain
\begin{equation}
\lim_{t\rightarrow\infty}\left\langle\frac{d\omega}{dt}\right\rangle=\frac{\pi
c\omega_{p0}^{2}\omega_{\phi}^{2}k_{\phi}\omega_{pl}\kappa^{2}}{8}\frac{\partial}{\partial\omega_{0}}\left[\frac{1}{\omega_{0}^{2}}\delta\left(\omega_{0}-\omega_{\phi}\right)\right].
\end{equation}
Inserting Eq.(21) into Eq.(19) the average rate of energy change of the electromagnetic field in plasma wave is expressed by
\begin{equation}
\overline{\langle\dot{\omega}\rangle}=-\frac{\pi
c\omega_{p0}^{2}k_{\phi}\omega_{pl}\kappa^{2}}{8}\frac{\partial
N\left(\omega_{0}\right)}{\partial\omega_{0}}\Bigg|_{\omega_{0}=\omega_{\phi}}.
\end{equation}
Eq.(22) indicates a mechanism of energy exchange of the electromagnetic wave with plasma wave. The frequency $\omega_{\phi}$ can be regarded as the resonance frequency because the photon with this frequency has the velocity matching with the phase velocity of the plasma wave. From Eq.(22) we find that the energy exchange is determined by the derivative of the distribution function of photon number near $\omega_{\phi}$, and if the slope of this distribution function is steeper the energy exchange is more remarkable. That just means only the photons having the frequencies close to the resonance frequency are involved in this process of energy exchange with the plasma wave. From the preceding discussion it is clear that the energy exchange is not related to collisions between photons and electrons. Thus the mechanism of the energy exchange here is a collisionless resonance effect between photon system and the plasma wave, namely the photon Landau damping[24,25]. Photon Landau damping has some properties in common with electron Landau damping: Both are collective velocity resonance phenomena with plasma waves. However the differences between them are obvious. The energy exchange in the process of electron Landau damping is achieved through doing work to electrons by the electrostatic field. But photons are electrically neutral and not influenced by the electrostatic force. In fact the energy exchange in photon Landau damping is achieved through the effect of photon frequency shift in plasma wave that is related to the fluctuation of the electron number density.

The plasma wave is described by the Poisson equation
\begin{equation}
\nabla^{2}\phi=\frac{en_{0}f}{\varepsilon_{0}}=-\frac{dE}{dx},
\end{equation}
where $\phi$ and $E$ are the scalar potential and strength of the longitudinal electric field in plasma wave. The solution of Eq.(23) is
\begin{equation}
E=\frac{en_{0}\kappa}{\varepsilon_{0}k_{pl}}\sin\left(k_{pl}x-\omega_{pl}t\right)=\hat{E}\sin\left(k_{pl}x-\omega_{pl}t\right).
\end{equation}
Then the amplitude of plasma wave $\kappa$ can be written as $\kappa=(\varepsilon_{0}k_{pl}\hat{E})/(en_{0})$, and Eq.(22) is given by
\begin{equation}
\overline{\langle\dot{\omega}\rangle}=-\frac{\pi
c\varepsilon_{0}k_{\phi}\omega_{pl}k_{pl}^{2}\hat{E}^{2}}{8mn_{0}}\frac{\partial
N\left(\omega_{0}\right)}{\partial\omega_{0}}\Bigg|_{\omega_{0}=\omega_{\phi}}.
\end{equation}
If we only consider the effect of photon Landau damping between photon system and the plasma wave and neglect other interaction, the damping rate $\gamma$ can be worked out by means of the energy conservation
\begin{equation}
\frac{1}{2}\varepsilon_{0}\hat{E}^{2}\gamma+\hbar\overline{\langle\dot{\omega}\rangle}=0.
\end{equation}
Substituting Eq.(25) into (26) the damping rate is obtained as
\begin{equation}
\gamma=\frac{\pi c\hbar k_{\phi}\omega_{pl}k_{pl}^{2}}{4mn_{0}}\frac{\partial N\left(\omega_{0}\right)}{\partial\omega_{0}}\Bigg|_{\omega_{0}=\omega_{\phi}}.
\end{equation}

We consider the electromagnetic wave as a time-Gaussian laser pulse, the initial photon number distribution in frequency domain space is expressed by
\begin{equation}
N\left(\omega_{0}\right)d\omega_{0}=N_{0}\frac{\tau\omega_{c}}{\sqrt{2\pi}\omega_{0}}\exp\left[-\frac{1}{2}\left(\omega_{0}-\omega_{c}\right)^{2}s^{2}\right]d\omega_{0},
\end{equation}
where $N_{0}$ is a normalization constant depending on the intensity of the initial laser pulse and $\omega_{c}$ is the center frequency and $s$ is the duration time. Substituting Eq.(28) into (27) the damping rate of time-Gaussian laser pulse to the plasma wave is derived as
\begin{equation}
\gamma=\sqrt{\frac{\pi}{2}}\frac{\hbar\tau\omega_{pl}k_{pl}^{2}\omega_{c}N_{0}}{4mn_{0}}\frac{ck_{\phi}}{\omega_{\phi}}\left(\left(\omega_{c}-\omega_{\phi}\right)s^{2}
-\frac{1}{\omega_{\phi}}\right)\exp\left[-\frac{1}{2}\left(\omega_{c}-\omega_{\phi}\right)^{2}s^{2}\right].
\end{equation}
In Ref.26 the same damping rate is obtained by using the kinetic theory. If the phase velocity of the plasma wave satisfies the condition $v_{\phi}<v_{c}-(c\omega_{p}^{2})/(\omega_{c}^{4}s^{2})$, where $v_{c}=c\sqrt{1-\omega_{p}^{2}/\omega_{c}^{2}}$, we have the damping rate $\gamma>0$ for the time-Gaussian laser pulse.

\section{Photon Evolution in Plasma Wave with Arbitrary Amplitude}
Now we will discuss the situation of the plasma wave with arbitrary amplitude. It is more convenient to perform a coordinate transformation from the laboratory frame to a new frame: $\xi=x-v_{\phi}t$, $\tau=t$. In this new frame the shape of plasma wave has a simpler expression as $f=f(k_{pl}\xi)$, and the photon dynamical equations are derived from Eq.(6) and (7) as
\begin{equation}
\frac{d\xi}{d\tau}=v-v_{\phi}=c\sqrt{1-\frac{\omega_{p}^{2}}{\omega^{2}}}-v_{\phi},
\end{equation}
\begin{equation}
\frac{d\omega}{d\tau}=-\frac{\omega_{p0}^{2}v_{\phi}}{2\omega}\frac{\partial f(k_{pl}\xi)}{\partial\xi}.
\end{equation}
In Eq.(31) we have canceled the photon momentum $k$ by using of the dispersion relation of electromagnetic wave in plasmas. Combining Eqs.(30) and (31) the time $\tau$ can be canceled and the evolution equation of photon in $\xi-\omega$ space is given by
\begin{equation}
\frac{dW}{df}=-\frac{\beta_{\phi}}{\sqrt{1-(1+f)/W}-\beta_{\phi}},
\end{equation}
where $\beta_{\phi}=v_{\phi}/c$ and $W=\omega^{2}/\omega_{p0}^{2}$. Here we also call the $\xi-\omega$ space as the phase space for convenience. The solution of Eq.(32) is
\begin{equation}
\sqrt{\alpha}\left(\sqrt{W}-\beta_{\phi}\sqrt{W-f-1}\right)=1,
\end{equation}
where $\alpha$ is a constant determined uniquely by the initial condition
\begin{equation}
\alpha=\frac{1}{\left(\sqrt{W_{0}}-\beta_{\phi}\sqrt{W_{0}-f_{0}-1}\right)^{2}}.
\end{equation}
In Eq.(34) $W_{0}$ depends on the initial photon frequency and $f_{0}$ is determined by the initial position of the photon injected into the plasma wave as $f_{0}=f(k_{pl}\xi_{0})$. Eq.(33) indicates that $\left(\sqrt{W}-\beta_{\phi}\sqrt{W-f-1}\right)$ is an invariant as long as the initial conditions are given. From Eq.(33) the photon frequency is obtained as
\begin{equation}
\frac{\omega}{\omega_{p0}}=\frac{\gamma_{\phi}^{2}}{\sqrt{\alpha}}\left(1\pm\beta_{\phi}\sqrt{1-\frac{\alpha}{\gamma_{\phi}^{2}}(1+f)}\right),
\end{equation}
which describes the evolution trajectory of a photon in phase space.

In Eq.(35) the notations ``$\pm$'' denotes the two different branches of the trajectory in phase space, which represents the evolutions of photons with two different kinds of initial conditions. The sign ``$+$'' stands for the upper branch and ``$-$'' denotes the lower branch. We also suppose the plasma wave has the shape of cosine, $f(\xi)=\kappa\cos(k_{pl}\xi)$, for simplicity. The photon evolution trajectory in phase space expressed by Eq.(35) is illustrated in Fig.1, and the similar result was numerically obtained in Ref.22.
\begin{figure}[h]
\centering
\includegraphics[width=4in]{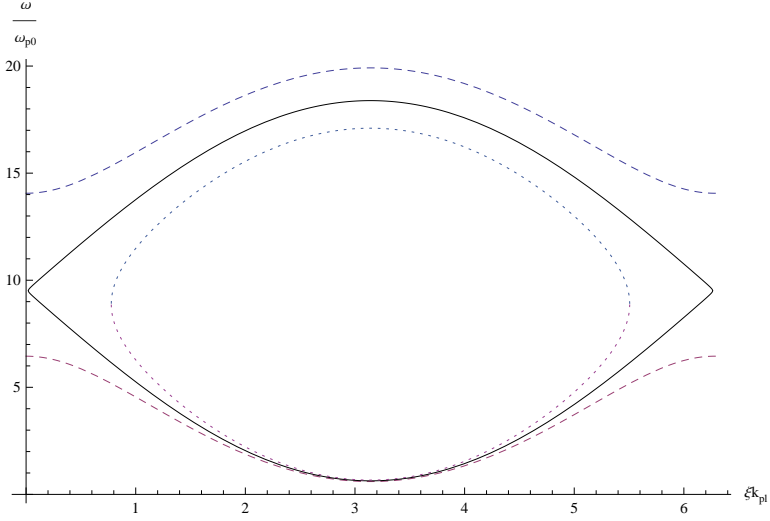}
\caption{The photon evolution trajectory in phase space. The dotted line denotes the evolution trajectory of trapped photon which is confined to one wavelength area of the plasma wave and oscillates between upper and lower branches. The dashed line denotes the evolution trajectory of untrapped photon, and these photons can propagate through different periods of plasma wave. The solid line is the separatrix between the trapped and untrapped photon trajectories.}
\end{figure}
According to the different initial conditions the evolution trajectory can be divided into two types: The first one is displayed as the dotted line in Fig.1, where the upper and lower branches intersect with each other to form a closed trajectory. The photons evolving in this trajectory will be confined to one wavelength area of the plasma wave and oscillate between the upper and lower branches. This type of photon is known as the trapped photon by the plasma wave. The second type is displayed as the dashed line, where the upper and lower branches keep separated consistently. In this situation the photon evolves only on one branch and propagates through different periods of the plasma wave, so it is obviously not trapped by the plasma wave. If the photon is on the upper branch of the trajectory at the initial time, it will always evolve on the upper branch with the motion velocity faster than the phase velocity of the plasma wave. Whereas the photon will evolve on the lower branch and propagate backward with respect to the plasma wave if its initial position is on the lower branch in phase space. In Fig.1 the solid line denotes the separatrix between trajectories of trapped and untrapped photons. Whether the photon can be trapped is determined by initial conditions. We have known that the trapped photon has the closed evolution trajectory, that is to say the factor under the square root sign in Eq.(35)
\begin{equation}
1-\frac{\alpha}{\gamma_{\phi}^{2}} (1+\kappa\cos(k_{pl}\xi))=0
\end{equation}
is solvable. Then we get the trapped condition for photons in plasma wave as
\begin{equation}
\frac{\gamma_{\phi}^{2}}{1+\kappa}\leq\alpha\leq\frac{\gamma_{\phi}^{2}}{1-\kappa}.
\end{equation}
All the photons with initial conditions satisfying inequality (37) can be trapped by the plasma wave. The trapped photons have the velocity matching with the phase velocity of plasma wave, and can exchange energy effectively with the plasma wave.

Let us discuss the motion equation of photons in coordinate space. The integral of dynamical equation (30) gives
\begin{equation}
\int_{\xi_{0}}^{\xi}\frac{d\xi'}{\sqrt{1-\left(1+f(\xi')\right)/W(\xi')}-\beta_{\phi}}=c\tau.
\end{equation}
Supposing that the plasma wave has the shape of cosine just as we discussed before and inserting Eqs.(33) and (35) into Eq.(38), the motion equation is given by
\begin{equation}
c\tau=\beta_{\phi}\gamma_{\phi}^{2}(\xi-\xi_{0})\pm\frac{\gamma_{\phi}^{2}}{k_{pl}}\int_{k_{pl}\xi_{0}}^{k_{pl}\xi}\frac{dx}{\sqrt{\left(1-\frac{\alpha}
{\gamma_{\phi}^{2}}-\frac{\alpha\kappa}{\gamma_{\phi}^{2}}\right)+\frac{2\alpha\kappa}{\gamma_{\phi}^{2}}\sin^{2}\frac{x}{2}}}.
\end{equation}
The sign ``$+$'' in Eq.(39) denotes the motion trajectory of the photon evolving on the upper branch, and ``$-$'' represents the trajectory of the photon evolving on the lower branch.

We will analyze the motion equation of trapped photon first. The maximum and minimum frequencies of the trapped photon in plasma wave can be obtained from Eq.(35) and given by
\begin{equation}
\omega_{max}=\frac{\gamma_{\phi}^{2}\omega_{p0}}{\sqrt{\alpha}}\left(1+\beta_{\phi}\sqrt{1-\frac{\alpha}{\gamma_{\phi}^{2}}(1-\kappa)}\right),
~~\omega_{min}=\frac{\gamma_{\phi}^{2}\omega_{p0}}{\sqrt{\alpha}}\left(1-\beta_{\phi}\sqrt{1-\frac{\alpha}{\gamma_{\phi}^{2}}(1-\kappa)}\right).
\end{equation}
The initial conditions of the trapped photon satisfy inequality (37), which gives $1-\alpha/\gamma_{\phi}^{2}-\alpha\kappa/\gamma_{\phi}^{2}<0$, and then the motion equation (39) is simplified into
\begin{equation}
c\tau=\beta_{\phi}\gamma_{\phi}^{2}(\xi-\xi_{0})\pm\frac{i2\gamma_{\phi}^{2}}{k_{pl}\sqrt{\frac{\alpha\kappa}{\gamma_{\phi}^{2}}
+\frac{\alpha}{\gamma_{\phi}^{2}}-1}}\int_{\frac{k_{pl}\xi_{0}}{2}}^{\frac{k_{pl}\xi}{2}}\frac{d\theta}{\sqrt{1-k^{2}\sin^{2}\theta}},
\end{equation}
with
\begin{equation}
k^{2}=\frac{(2\alpha\kappa/\gamma_{\phi}^{2})}{(\alpha\kappa/\gamma_{\phi}^{2})+(\alpha/\gamma_{\phi}^{2})-1}>0.
\end{equation}
The integral function in Eq.(41) can be expressed by the elliptic integral of the first kind $F(k_{pl}\xi/2,k)$. For simplicity we only discuss the motion equation in the first period, the evolution after the first period can be worked out analogously. Assuming that the photon is on the upper branch of the trajectory in phase space at the initial time, the motion trajectory before the photon going into the lower branch is derived by using Eq.(41) as
\begin{equation}
c\tau=\beta_{\phi}\gamma_{\phi}^{2}(\xi-\xi_{0})-\frac{2\gamma_{\phi}^{2}}{k_{pl}\sqrt{\frac{\alpha\kappa}{\gamma_{\phi}^{2}}
+\frac{\alpha}{\gamma_{\phi}^{2}}-1}}\mathrm{Im}\left[F(k_{pl}\xi/2,k)-F(k_{pl}\xi_{0}/2,k)\right].
\end{equation}
And then, the photon reaches the lower branch and the motion equation is given by
\begin{equation}
c\tau=\beta_{\phi}\gamma_{\phi}^{2}(\xi-\xi_{0})-\frac{2\gamma_{\phi}^{2}}{k_{pl}\sqrt{\frac{\alpha\kappa}{\gamma_{\phi}^{2}}
+\frac{\alpha}{\gamma_{\phi}^{2}}-1}}\mathrm{Im}\left[2F(k_{pl}\xi_{m}/2,k)-F(k_{pl}\xi_{0}/2,k)-F(k_{pl}\xi/2,k)\right],
\end{equation}
where $\xi_{m}$ determined by
\begin{equation}
1-\frac{\alpha}{\gamma_{\phi}^{2}}\left(1+\kappa\cos(k_{pl}\xi_{m})\right)=0
\end{equation}
is the maximum position the photon can reach. At last the photon returns to the upper branch and finishes the evolution in a completed cycle. If the photon is on the lower branch of the trajectory in phase space at the initial time, we can analyze the motion equation by using the similar method, and the motion trajectory of this kind of photon is illustrated in Fig.2.
\begin{figure}[h]
\centering
\includegraphics[width=4in]{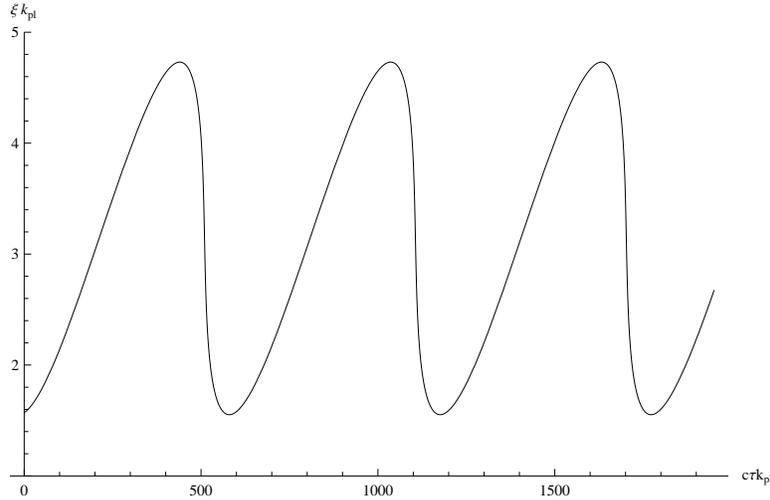}
\caption{The motion trajectory of trapped photon in coordinate space. At the initial time the photon is on the lower branch of the trajectory in phase space. This kind of photon is confined to the area shorter than one wavelength of the plasma wave and oscillates between upper and lower branches, so the motion trajectory in coordinate space is also oscillatory as displayed in the figure.}
\end{figure}
It is shown that the trajectory is oscillatory, and the photon is confined to the area shorter than one wavelength of the plasma wave. The velocity of the photon is close to the phase velocity of the plasma wave.

The situation of the untrapped photon is somewhat simpler. If the photon is on the upper branch of the trajectory in phase space at the beginning, it always propagate on the upper branch, and the maximum and minimum frequencies of the photon in plasma wave are given by
\begin{equation}
\omega_{max}=\frac{\gamma_{\phi}^{2}\omega_{p0}}{\sqrt{\alpha}}\left(1+\beta_{\phi}\sqrt{1-\frac{\alpha}{\gamma_{\phi}^{2}}(1-\kappa)}\right),
~~\omega_{min}=\frac{\gamma_{\phi}^{2}\omega_{p0}}{\sqrt{\alpha}}\left(1+\beta_{\phi}\sqrt{1-\frac{\alpha}{\gamma_{\phi}^{2}}(1+\kappa)}\right).
\end{equation}
Inequality (37) is not satisfied now and we have $1-\alpha/\gamma_{\phi}^{2}-\alpha\kappa/\gamma_{\phi}^{2}>0$, so the motion equation (39) gives
\begin{equation}
c\tau=\beta_{\phi}\gamma_{\phi}^{2}(\xi-\xi_{0})+\frac{2\gamma_{\phi}^{2}}{k_{pl}\sqrt{1-\frac{\alpha}{\gamma_{\phi}^{2}}-
\frac{\alpha\kappa}{\gamma_{\phi}^{2}}}}\mathrm{Re}\left[F(k_{pl}\xi/2,ik)-F(k_{pl}\xi_{0}/2,ik)\right],
\end{equation}
with
\begin{equation}
k^{2}=\frac{(2\alpha\kappa/\gamma_{\phi}^{2})}{1-(\alpha/\gamma_{\phi}^{2})-(\alpha\kappa/\gamma_{\phi}^{2})}>0.
\end{equation}
Eq.(47) is the motion equation of the untrapped photon with the initial position on the upper branch. In Fig.3 the motion trajectory of this type of photon in coordinate space is displayed.
\begin{figure}[h]
\centering
\includegraphics[width=4in]{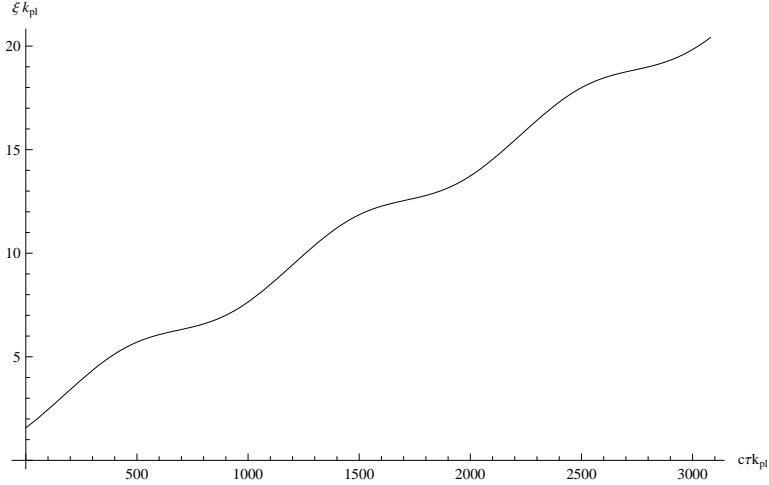}
\caption{The motion trajectory of trapped photon with the initial position on the upper branch in phase space. In this situation the photon always propagate with the velocity faster than the phase velocity of the plasma wave, so it travels forward with respect to the plasma wave.}
\end{figure}
The photon velocity is always faster than the phase velocity of the plasma wave, so the photon propagates forward with respect to the plasma wave and can enter into different periods of the plasma wave. If the photon is on the lower branch of the trajectory in phase space at the initial time, it will always propagate on the lower branch, and the maximum and minimum frequencies of the photon in plasma wave are obtained by using Eq.(35) as
\begin{equation}
\omega_{max}=\frac{\gamma_{\phi}^{2}\omega_{p0}}{\sqrt{\alpha}}\left(1-\beta_{\phi}\sqrt{1-\frac{\alpha}{\gamma_{\phi}^{2}}
(1+\kappa)}\right),~~\omega_{min}=\frac{\gamma_{\phi}^{2}\omega_{p0}}{\sqrt{\alpha}}\left(1-\beta_{\phi}\sqrt{1-\frac{\alpha}{\gamma_{\phi}^{2}}
(1-\kappa)}\right).
\end{equation}
The motion equation of this kind of photon is given by
\begin{equation}
c\tau=\beta_{\phi}\gamma_{\phi}^{2}(\xi-\xi_{0})-\frac{2\gamma_{\phi}^{2}}{k_{pl}\sqrt{1-\frac{\alpha}{\gamma_{\phi}^{2}}-
\frac{\alpha\kappa}{\gamma_{\phi}^{2}}}}\mathrm{Re}\left[F(k_{pl}\xi/2,ik)-F(k_{pl}\xi_{0}/2,ik)\right].
\end{equation}
We illustrate the motion trajectory in Fig.4. The velocity of this type of photon is always slower than the phase velocity of the plasma wave, and the photon moves backward with respect to the plasma wave.
\begin{figure}[h]
\centering
\includegraphics[width=4in]{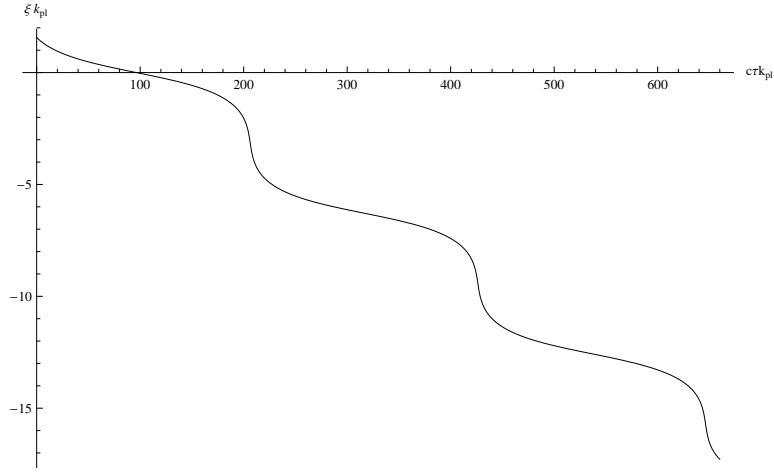}
\caption{The motion trajectory of trapped photon with the initial position on the lower branch in phase space. This kind of photon always moves with the velocity slower than the phase velocity of the plasma wave, so it travels backward with respect to the plasma wave.}
\end{figure}
In Fig.5 we show the motion trajectory of the photon evolving on the separatrix in one complete cycle. The photon is just restricted to one wavelength area of the plasma wave and oscillates between two neighbor peaks of the plasma wave.
\begin{figure}[h]
\centering
\includegraphics[width=4in]{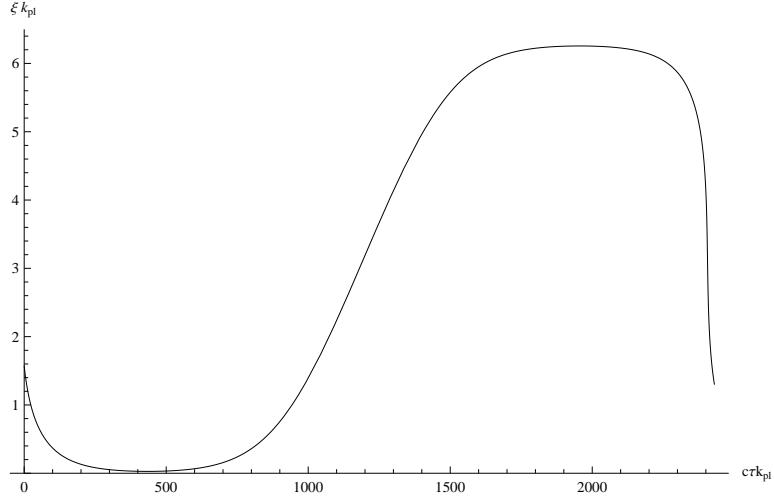}
\caption{The motion trajectory of the photon evolving on the separatrix in one complete cycle.}
\end{figure}

At last let us discuss a simple case in the limit of $v_{\phi}\rightarrow c$. The evolution equation of a photon in phase space, Eq.(33), is simplified as
\begin{equation}
W=\frac{1}{4\alpha}\left[\alpha(1+f)+1\right]^{2}
\end{equation}
in this situation, which gives
\begin{equation}
\omega=\frac{\omega_{p0}}{2\sqrt{\alpha}}\left(\alpha\kappa\cos(k_{pl}\xi)+\alpha+1\right).
\end{equation}
Considering the region of underdense plasma $\omega\gg\omega_{p0}$, i.e., $W\gg1$, both $1/\alpha$ and $(1+f)/W$ are small quantities, and the motion equation (38) can be expanded in series and calculated to the first order, which gives
\begin{equation}
\alpha\kappa\sin(k_{pl}\xi)+(\alpha+2)k_{pl}\xi+2c\tau k_{pl}=\alpha\kappa\sin(k_{pl}\xi_{0})+(\alpha+2)k_{pl}\xi_{0}.
\end{equation}
Apparently, in this limit the photon cannot be trapped by the plasma wave because its velocity is slower than $c$ all the time, and it will propagate backward with respect to the plasma wave.

\section{Conclusion and Discussion}
In this paper some dynamical behaviors of photons and collective effect of photon system in plasma wave are described in the unified frame of the photon dynamical theory. We treat the plasma wave as a special background for photons, and the refractive index of the plasma is equivalent to an external potential field which determines the dynamical evolution of photons. The dynamical equations of photons are constructed in Hamiltonian formulation in the approximation of geometrical optics. We solve the dynamical equations and analyze some motion behaviors of photons in plasma wave, including photon acceleration and trapping effects, the evolution trajectory in phase space and photon Landau damping to the plasma wave. In fact the photon trapping effect in plasma wave is related to the photon acceleration. Only when the photon velocity is close to the phase velocity of plasma wave, the trapping phenomenon occurs, which is analogous to the electron trapping effect in a plasma wave. Whether a photon can be accelerated to a velocity that matches with the phase velocity of a given plasma wave is determined by the initial conditions. The trapping condition and possibility of photons in a given plasma wave are analysed in detail in this paper. In a small-amplitude plasma wave, the evolution behavior of a single photon is extended to the entire electromagnetic field to calculate the collisionless Landau damping effect originated from the electromagnetic field. When the plasma wave gets a relativistic phase velocity, for example a plasma wave driven by a laser pulse, some photons of the electromagnetic field can propagate with velocities very close to the phase velocity of the plasma wave and be trapped by the plasma wave, and consequently the velocity resonance with plasma wave occurs. The energy can be exchanged effectively between the electromagnetic field and the plasma wave in this resonance process. If we consider the electromagnetic field as a laser pulse, due to the narrow spectrum width most photons in the pulse may resonate with the plasma wave to produce the photon Landau damping effect.

Theoretically, if we know the dynamical evolutions of photons in plasma waves, the evolution properties of the laser pulse can be obtained by making statistical summation of all photon behaviors, in which the keypoint is the weight factor of photons with different modes in entire pulse, i.e., photon number distribution. In a small-amplitude plasma wave the modulation on photon motion by the plasma wave is so weak that photon number distribution can be considered as a invariant, so we can use the initial pulse distribution as the statistical weight and keep it unchanged. However in a large-amplitude plasma wave the photon motion is significantly modulated and the photon number distribution evolves to obey the kinetic equation. Consequently, it is also an important problem to solve the photon kinetic equation in a large-amplitude plasma wave so as to obtain the time-dependent photon number distribution in phase space[32]. The photon dynamics plays an important role in the research of the interaction of electromagnetic field with plasma wave based on the corpuscular theory of light. By combining the kinetic theory of photon system and the single photon dynamics, the evolution of an electromagnetic field in a given plasma wave can be well-described in a different method from the Maxwell wave theory.

\begin{acknowledgments}
This work was partly supported by the Shanghai Leading Academic Discipline Project (Project No. S30105), and Shanghai Research Foundation (Grant No. 07dz22020).
\end{acknowledgments}

\end{document}